\documentclass[aps,showpacs,twocolumn]{revtex4}

\usepackage{graphicx}
\usepackage{epsfig}
\usepackage{amsfonts}
\usepackage{amsmath,amssymb}

\begin{document}

\title{Exact solution of the spin-isospin proton-neutron pairing Hamiltonian}

\author{ S. Lerma H.$^{1}$, B. Errea$^{1}$,  J. Dukelsky$^{1}$,  and
        W. Satu{\l}a$^{2}$}
\address{$^{1}$ Instituto de Estructura de la Materia, CSIC, Serrano 123, 28006 Madrid, Spain. \\
         $^{2}$ Institute of Theoretical Physics, University of Warsaw, ul. Hoza 69, 00-681 Warsaw, Poland.}

\date{\today}

\begin{abstract}
The exact solution of  proton-neutron isoscalar-isovector ($T$=0,1) pairing Hamiltonian with non-degenerate
single-particle orbits  and equal pairing strengths ($g_{T=1}$=$g_{T=0}$) is presented for the first time. The
Hamiltonian is a particular case of a family of integrable SO(8) Richardson-Gaudin (RG) models. The exact solution
of the $T$=0,1 pairing Hamiltonian is reduced to a problem of 4 sets of coupled non linear equations that
determine the spectral parameters of the complete set of eigenstates. The microscopic structure of individual
eigenstates is analyzed in terms of evolution of the spectral parameters in the complex plane for system of $A$=80
nucleons. The spectroscopic trends of the exact solutions are discussed in terms of generalized rotations in
isospace.
\end{abstract}

 \maketitle

The exactly solvable models
introduced by Richardson \cite{Rich1} and
by Gaudin \cite{Gaud1} belong nowadays to classic theoretical
tools in mesoscopic physics. Indeed, these
models  based on the rank 1 SU(2) algebra for fermions or the SU(1,1) algebra
for bosons were applied to a large variety of quantum many-body systems
including the atomic nucleus, superconducting grains, cold atomic gases,
etc., see review article~\cite{Duk} and refs. therein.
Recently, we have extended the Richardson-Gaudin (RG)
models to the rank 2 algebras: SO(5)
(isovector pairing \cite{o5}), SO(3,2) (F spin 1 boson pairing \cite{so32}),
and SU(3) (interacting three level atoms \cite{su3}).

In this letter we will derive for the first time the exact solution
for the rank 4 SO(8) RG integrable model
with non-degenerate single-particle ({\it sp\/}) spectrum and arbitrary
degeneracies.  As a particular realization of the rank 4 SO(8) RG model
we will consider the nuclear isoscalar-isovector ($T$=0,1)
pairing Hamiltonian introduced for a single degenerate shell
in Ref. \cite{Flowers} and further developed in \cite{pang,dussel}.
We will solve the model for a realistic case of $A$=80 nucleons moving in
fourfold degenerated equidistant $sp$ spectrum working
in $T$=0,1 pair representation of the SO(8) algebra.
It should be mentioned that other
representations like the Ginnocchio model \cite{Gin} can lead to
interesting exactly solvable models in nuclear structure as well as to models
of spin $3/2$ cold atoms \cite{3/2}.

The study of proton-neutron (p-n) pairing has gained a renewed interest due to the new generation of
radioactive-beam facilities that will open the access to proton-rich nuclei close to the $N$=$Z$ line. In spite of
vigorous activity in this field, see \cite{Van} and refs. therein, the fundamental questions concerning the basic
building blocks and experimental fingerprints of the p-n pairing are still a matter of debate. So are the
theoretical problems concerning generalization of well established nuclear pairing models to include p-n pairing,
proper treatment of isospin degree of freedom or $\alpha$-like clustering. All these problems set clear motivation
for realistic exact-model studies of the p-n pairing undertaken in this work.

 Let us begin our derivation by introducing  the 28 generators of the SO(8) algebra~\cite{Flowers}:
    three ($T$=1,$S$=0)  and  three   ($T$=0, $S$=1) pair creators,
together with their respective annihilation operators:
$\mathbf{P}_{\tau i}^\dagger=
\sqrt{\frac{2l_i+1}{2}}[\mathbf{a}^\dagger_{l_{i}}\mathbf{a}^\dagger_{l_i}]^{010}_{0\tau 0}$,
$\mathbf{D}_{\sigma i
}^\dagger=\sqrt{\frac{2l_i+1}{2}}[\mathbf{a}^\dagger_{l_i}\mathbf{a}^\dagger_{l_i}]^{001}_{00\sigma}$ ,
$\mathbf{P}_{\tau i}=(\mathbf{P}_{\tau i}^\dagger )^\dagger$,
and $\mathbf{D}_{\sigma
i }=(\mathbf{D}_{\sigma i}^\dagger)^\dagger$,
where the triads  in the couplings represent, respectively, angular momentum, isospin and spin. The fermionic
operators $\mathbf{a}^\dagger_{l_i m\tau\sigma}$ create a particle in the orbit $l_i$ with projection $m$, isospin
$\tau$ and spin $\sigma$.
 The SO(8) algebra is completed by the $16$ particle-hole operators:
$\mathbf{C}_{\tau_1\sigma_1,\tau_2\sigma_2; i}\equiv \sum_m \mathbf{a}^\dagger_{l_i m
\tau_1\sigma_1}\mathbf{a}_{l_i m\tau_2\sigma_2} $. These 16 operators close an U(4) subalgebra of SO(8) and  include
the number operators for the four different nucleon types  in the orbit $l_i$: $\mathbf{N}_{\tau\sigma
i}=\mathbf{C}_{\tau\sigma, \tau\sigma; i}$.

The general procedure for solving RG models for arbitrary simple Lie algebras has been developed in references
\cite{Ushve, AFS}. For each algebra it is possible to derive a set of $L$ quadratic  integrals of motion
 defining the integrable model. It is also possible to derive the complete set of common eigenstates and eigenvalues,
which constitute the exact solution of the model. Here $L$ is the number of copies of the algebra that we
associate with the number of orbits $l_i$. For simplicity we will be concerned here with a particular Hamiltonian,
the $T=0,1$ proton-neutron pairing Hamiltonian that arises as a particular linear combinations of the integrals of
motion of the SO(8) RG model:
\begin{equation}
\mathbf{H}=\sum_i^L \epsilon_i\mathbf{N}_i-g\sum_{ii'}^L
\sum_{\mu}\left(\mathbf{P}^\dagger_{\mu i}\mathbf{P}_{\mu
{i'}} +
 \mathbf{D}^\dagger_{\mu i}\mathbf{D}_{\mu
 {i'}}\right),
 \label{hamilton}
\end{equation}
where $\mathbf{N}_i$ is the number operator of the orbit $l_i$.

We would like to emphasize here that the Hamiltonian (\ref{hamilton}) has equal strength for $T$=1 and $T$=0
pairing. As a consequence, there is a conserved U(4) symmetry defined by the generators
  $\mathbf{C}_{\tau_1\sigma_1, \tau_2\sigma_2}\equiv\sum_{i}^L\mathbf{C}_{\tau_1\sigma_1,\tau_2\sigma_2; i}$.
Therefore,  the   eigenstates  are organized in degenerated U(4) Wigner multiplets.   For a given number of
nucleons $A$, these  U(4) multiplets can be classified using Young tableaux. Each multiplet  is
defined by a partition of $A$ in  4 numbers,  $[h_1 h_2 h_3 h_4]$, constrained by: $ \sum_i (2l_i+1)\geq h_1\geq
h_2\geq h_3\geq h_4\geq 0$.  The labels $h_i$ are related to the number of particles in the total
U(4) Lowest Weight State (LWS). For instance, if we relabel the U(4) operators according to the rule $1\equiv
n\!\!\downarrow$, $2\equiv n\!\!\uparrow$, $3\equiv p\!\!\downarrow$ and $4\equiv p\!\!\uparrow$, the  U(4) LWS
can be defined as the state which  satisfies: $\mathbf{C}_{\alpha, \beta}|LWS\rangle=0$, $\forall \alpha<\beta$.
For this choice of LWS, the corresponding U(4) Young tableau is given by: $[N_{n\downarrow} N_{n\uparrow}
N_{p\downarrow} N_{p\uparrow}]$. The spin and isospin of this LWS  are simply
$2S\!\!=\!\!N_{n\downarrow}\!\!+\!\!N_{n\downarrow}\!\!-\!\!
N_{n\uparrow}\!\!-\!\!N_{n\uparrow}$ and $2T\!\!=\!\!N_{n\downarrow}\!\!+\!\! N_{n\uparrow}\!\!-\!\!N_{p\downarrow}\!\!-\!\! N_{p\uparrow}$.

As stated above, the eigenvalues of Hamiltonian  (\ref{hamilton}) can be
derived from the exact solution of the SO(8) RG model. They are:
 \begin{equation}
E=\sum_{\alpha}^{M_1}
e_{\alpha}+\sum_{i=1}^L
\epsilon_i u_i.
\end{equation}
where $u_i$ is the  seniority of level $i$, i.e., the number of particles in level $i$ not coupled in $T$=1 or $T$=0
pairs.   The parameters $e_\alpha$ satisfy  the generalized Richardson equations:
\begin{eqnarray}
\sum_{\alpha'(\not=\alpha)}^{M_1}\frac{2}{e_{\alpha'}-e_{\alpha}}-\sum_{\alpha'}^{M_2}\frac{1}{\omega_{\alpha'}-e_{\alpha}}&\nonumber\\
-\sum_{i}^L\frac{(2l_i+1)-h_{1; i}-h_{2; i}
}{2 \epsilon_i-e_{\alpha}}+\frac{1}{g}&=0
\nonumber\\
\sum_{\alpha'(\not=\alpha)}^{M_2}\frac{2}{\omega_{\alpha'}-\omega_{\alpha}}
-\sum_{\alpha'}^{M_1}\frac{1}{e_{\alpha'}-\omega_{\alpha}}-\sum_{\alpha'}^{M_3}\frac{1}{\eta_{\alpha'}-\omega_{\alpha}}&\label{reso8}\\
-\sum_{\alpha'}^{M_4}\frac{1}{\gamma_{\alpha'}-\omega_{\alpha}}
+\sum_i^L\frac{ h_{3; i}-h_{2; i}}{2\epsilon_i-\omega_\alpha}&=0
\nonumber\\
\sum_{\alpha'(\not=\alpha)}^{M_3}\frac{2}{\eta_{\alpha'}-\eta_{\alpha}}
-\sum_{\alpha'}^{M_2}\frac{1}{\omega_{\alpha'}-\eta_{\alpha}}+\sum_i^L\frac{h_{4; i}-h_{3; i}}{2\epsilon_i-\eta_\alpha}&=0 \nonumber\\
\sum_{\alpha('\not=\alpha)}^{M_4}\frac{2}{\gamma_{\alpha'}-\gamma_{\alpha}}
-\sum_{\alpha'}^{M_2}\frac{1}{\omega_{\alpha'}-\gamma_{\alpha}}+\sum_i^L\frac{h_{2; i}-h_{1; i}}{2\epsilon_i-\gamma_\alpha}&=0\nonumber,
\end{eqnarray}
where $[h_{1; i} h_{2; i} h_{3; i} h_{4; i}]$ is the Young tableau of the
  reduced U(4) irrep  defined by the unpaired particles in the
$i$-th orbit. These U(4) labels are constrained by the conditions: $2l_i+1\geq h_{1; i}\geq h_{2; i}\geq h_{3;
i}\geq h_{4; i}\geq 0$. In terms of these labels the seniority of  level  $i$ is  $u_i=\sum_{k} h_{k; i}$. The
rank of the RG models defines the number of different sets of spectral parameters. SO(8) is a rank 4 algebra,
hence there are 4 sets of spectral parameters. The number of spectral parameters in each set is determined by the
reduced labels and those of the total U(4) Wigner multiplet: $M_1=(A-u)/2$, $
 M_2=h_{3}+h_{4}-\sum_i (h_{3;i}+h_{4; i})$,
 $M_3= h_{4}-\sum_{i} h_{4;i}$, and
$M_4=(A-2 h_1)/2-(u-\sum_i 2h_{1; i})/2$,
with $u=\sum_i u_{i}$.

 The first set of spectral parameters comprises
 the usual pair energies $e_\alpha$ of the SO(8) algebra.
 The other three sets, composed by the spectral parameters $\omega_\alpha$,
$\eta_\alpha$ and $ \gamma_\alpha$, are associated with the U(4) subalgebra of SO(8). While the eigenvalues depend
only on the parameters $e_\alpha$, the corresponding eigenfunctions are determined by the parameters of the four
sets. The complete set of solutions of the Richardson equations  defines a basis which spans completely the
Hilbert space of sates with the same U(4) Wigner quantum numbers $[h_1 h_2 h_3 h_4]$.

 Even though the set of non linear coupled equations (\ref{reso8}) seems to be extremely complex,
 we will show how it is possible to obtain numerical solutions within a non trivial example of $A$=$4 n$
 nucleons with $n$ a positive
  integer moving in a set of non-degenerate $l=0$ single particle orbits. Other cases could be handle following
  a similar procedure.
  Before describing the numerical strategy, it will be useful to consider the lowest energy LWS configurations for
 even and odd isospin in the $g$=0 limit.

For even $T$ the lowest $(A-2T)/4$ levels are filled with 4 nucleons, and the following $T$ levels with a pair of
neutrons. All particles are paired and the corresponding seniority quantum numbers are 0. In the case of $T$ odd,
the levels $(A-2T+2)/4$ and $(A+2T+2)/4$ have one unpaired nucleon ($u_{(A-2T+2)/4} = u_{(A+2T+2)/4}= 1$). This
state can be considered a p-h excitation that evolves to a two quasi-particle state in the superconducting phase.
As a consequence, the number of pair energies $e_{\alpha}$ is $M_1=A/2$ in the even $T$ case and $M_1=A/2-1$ in
the odd $T$ case. In this limit the pair energies take the values $e_\alpha=2\epsilon_i$ according to the same
pattern.

 In the weak coupling limit ($g\!\ll\! 1$) the Richardson equations (\ref{reso8}) decouple  into independent sets of
 equations, each one related to the single particle level partially or fully occupied in the $g=0$ limit.
 These equations can be solved analytically. The  4 sets of spectral parameters obtained in this way are used
 as  initial guess for an iterative procedure in which the coupling constant $g$ is increased
step by step using the previous solution as the initial the guess.

As is well known, the main obstacle in solving the Richardson equations even in the SU(2) case, is the appearance
of singularities at some critical values of the pairing strength due to crossings in the real axis of single
particle energies and spectral parameters \cite{Romb, Ese}. This problem is even worst in the SO(8) model having 4
sets of spectral parameters. In order to avoid these numerical instabilities we introduce an alternate imaginary
term in the $sp$-energies $\epsilon_i\rightarrow \epsilon_i + (-)^i \Delta \sqrt{-1}$, which breaks the time
reversal symmetry and moves the solutions of (\ref{reso8}) away from the real axis \cite{Dus}. The system is then
evolved from the initial guess at $g\! \ll\! 1$ to the desired value of $g$. At this point we begin a second
iterative process to set the imaginary term in the single particle energies to zero ($\Delta\rightarrow 0$). This
recursive procedure proved to be very efficient in solving the Richardson equations for arbitrary values of the
coupling constant and for all the states considered.

We will now demonstrate the ability of our procedure for solving large scale complex physical problems by
presenting a numerical example for a system of $A$=80 nucleons described by the Hamiltonian (\ref{hamilton}) and
moving in  a set of $L$=50 equidistant and fourfold degenerate levels ($\epsilon_i=(i-1) /2$, $l_i=0$ with
$i=1,...,50$). The advantage of the method is that it provides not only exact eigenstates (or {\it
spectroscopic\/} information) but also allows for intuitive, pictorial representation of the {\it microscopic\/}
structure of individual eigenstates. In the following we will briefly present both aspects of our model.

  \begin{figure}
 \includegraphics[height=.4\textheight, width=0.46\textwidth]
 {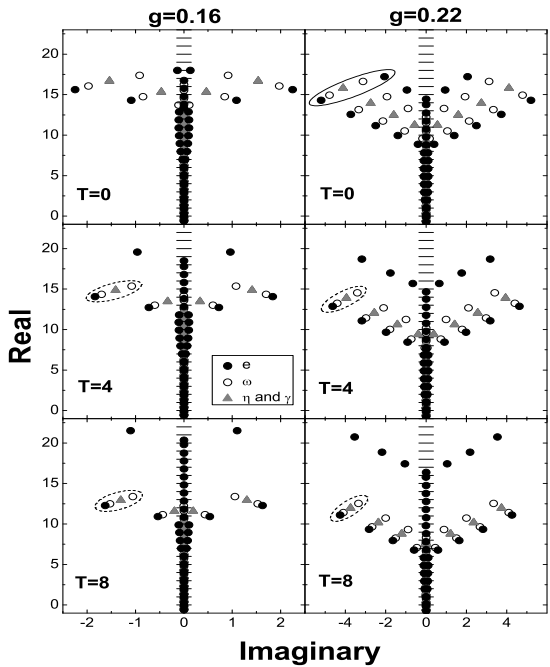}
\caption{Complex plane representation of the pair energies $e$ and wave function structure parameters $\omega$,
$\eta$ and $\gamma$, for the lowest-energy states with isospin $T$=0,4,8. The values of $g$ are indicated on the
top of the columns. Horizontal lines represent the $sp$-energies $2\epsilon_i=(i-1)$. Clusters representing alpha
quartets ($T$=0) and  collective  p-p pairs  ($T$=4 and $T$=8)   are indicated by full and dashed ellipses
respectively.
 }
\label{Fig2}
 \end{figure}

Let us start with a brief discussion of {\it microscopic\/} aspects of our solutions. Since the exact eigenstates
are fully determined by the spectral parameters ($e, \omega, \eta, \gamma$) the evolution of these parameters in
the complex plane allows for tracking structural changes of individual eigenstates as a function of model
parameters. This is demonstrated in Fig.~\ref{Fig2} where we show the spectral parameters in the complex plane for
three different isospin states and two values of the coupling constant representing weak $g=0.16$ and intermediate
$g=0.22$ pairing. As in the SU(2) RG model, the expansion of the pair energies $e$ into the complex plane
indicates the formation of correlated Cooper pairs~\cite{Tin}. Simple counting shows that, for the two $g$ values
considered here, the overall number of correlated pairs is, irrespective  of isospin, about $6$ ($15 \%$) and $14$
($35 \%$) respectively. The rest of the pairs, still attached to the $sp$ energies, remain almost uncorrelated.

The expansion of the spectral parameters $\omega$, $\eta$ and $\gamma$ in the complex plane follows the behavior
of the pair energies $e$, namely, they form parallel arcs to those of the pair energies and they arrange
themselves into various cluster-like structures, see Fig.~\ref{Fig2}. According to our choice of the LWS an
isolated complex parameter $e$ represents a collective $T$=1  n-n Cooper pair; a cluster of one $e$, two
$\omega$'s, one $\eta$, and one $\gamma$ represents a collective $T$=1 Cooper p-p pair, while a cluster of two
$e$'s, two $\omega$'s, one $\eta$, and one $\gamma$ represents a correlated alpha-like quartet. This allows for an
unambiguous interpretation of Fig.~\ref{Fig2}. The existence of correlated quartets is clearly visible in the
$T=0$ panels of the figure for strong enough $g$. With increasing $T$, these clusters break apart and the net
separation of the arcs of the pair energies increases. Moreover one of the arcs is formed by isolated pair
energies $e$ while the other one is constituted by pair energies forming clusters with two $\omega$'s, one $\eta$,
and one $\gamma$ spectral parameters. Physically, it implies a quenching of isoscalar pairing and the formation of
two separate conventional n-n and p-p pairing condensates.

Let us turn now to the {\it spectroscopic\/} consequences of the observed
microscopic processes along the nuclear symmetry energy (NSE) curve
$E(T)$. In the following we will show that these processes
form a systematic pattern which can be nicely interpreted in terms of
generalized rotations in isospace in the spirit of
iso-cranking model of Refs.~\cite{[Sat01]}.
Let us recall that according to that model the NSE splits into two
structurally different even-$T$ and
odd-$T$ branches of an iso-rotational band which can be conveniently
parametrized as: $E^{(e)}(T)  =  T(T+\lambda)/2J_T$ and $E^{(o)}(T)  =
T(T+\lambda)/2J_T + \Delta E_{exc}$ respectively. Here $J_T$ stands for
moment of inertia in the isospace (iso-MoI) while $\lambda$ determines
strength of the linear term $\sim T$ which is often called the Wigner energy.
The odd-$T$ sequence is shifted up
with respect to the even-$T$ branch by a  two-quasiparticle [2qp] excitation
energy $\Delta E_{exc}$.
Note a beautiful analogy to the spatial collective rotation
in even-even nuclei where odd-spin branch is also built upon 2qp
excitation.

\begin{figure}
\includegraphics[width=0.34\textwidth]
{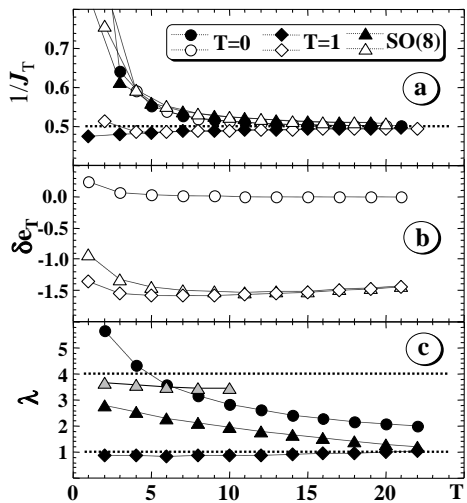}
\caption{
({\bf a}) Inverse of the iso-MoI ($1/J_T$), ({\bf b}) signature splitting
$\delta e_T$ and ({\bf c}) linear term enhancement factor $\lambda$
versus $T$ for pure $T$=0 model (circles), pure $T$=1 model (diamonds) and
the SO(8) model (triangles). Filled (open) symbols refer to
even-(odd-)$T$ branches of $E(T)$.
The calculations were done for $g=0.16$ except for gray triangles in
the lowest panel which mark the SO(8) solution for $g=0.22$.
}\label{fig3}
 \end{figure}

The calculated inverse of the iso-MoI ($1/J_T$) versus $T$, the primary characteristic of the iso-rotational
motion, is shown in Fig.~\ref{fig3}a. Apart from the SO(8) solution also two limiting SO(5) cases invoking only
isoscalar and only isovector pairing are depicted \cite{o5}. Note that in accordance to the iso-cranking model:
({\it i\/}) all curves converge to the $sp$ splitting $1/J_T\rightarrow \delta
\epsilon=1/2$~\cite{[Sat03],[Glo04]} ({\it ii\/}) the $T$=1 paring represents almost perfectly rigid rotation with
$1/J_T \approx \delta \epsilon$ irrespectively on $T$~\cite{[Sat01],[Glo04]} ({\it iii\/}) the SO(8) and $T$=0
pairing curves show  a characteristic reduction of the $J_T$ at low-$T$ due to isoscalar pairing collectivity
similar to the well recognized reduction of the spatial MoI caused by isovector superfluidity. The increase of
$J_T$ versus $T$ reflects disappearance of the isoscalar pairing collectivity caused by fast iso-rotation which
tends to recouple isoscalar (anti-parallel coupled isospins) pairs in analogy to the well known Coriolis
anti-pairing effect.

The quantity depicted in Fig.~\ref{fig3}b is directly related to average difference in pairing correlation energy
$E_{corr}\equiv \langle \hat H \rangle - E(T)$ ($\langle \rangle$ denotes the expectation value in the $g$=0
limit) between even and odd-$T$ branches. This quantity, known also as signature splitting, is defined at odd-$T$
as: $\delta e_T  \equiv (E^{(e)}(T+1) + E^{(e)}(T-1))/2 - E^{(o)}(T)$. At high-$T$, where $1/J_T \approx \delta
\epsilon$, signature splitting equals $\delta e_T = - \Delta E_{exc} + \delta \epsilon/2$ irrespective of
$\lambda$. In the case of pure $T$=0 pairing building up the isospin proceeds through the isoscalar pair breaking.
The correlation energy drops down and the solution goes smoothly over to the $sp$ limit where $\Delta E_{exc}
\rightarrow \delta \epsilon /2$. Consequently $\delta e_T \rightarrow 0$ as shown in Fig.~\ref{fig3}b. In the
$T$=1 pairing case the signature splitting is almost constant and equal $\delta e_T \approx -1.5$. In this case we
deal with rigid iso-rotation and odd and even-$T$ branches are shifted by a constant energy reflecting a
difference between correlation energies in odd (seniority two) and even-$T$ (seniority zero) states (2qp energy).
Finally, the SO(8) curve goes smoothly over to the $T$=1 case as the isoscalar pairing disappears with increasing
$T$.

Fig.~\ref{fig3}c shows the linear enhancement factor calculated as
$\lambda \equiv 2E_T^{(e)} \bar{J}_T / T - T$ where  $\bar{J}_T$
stands for mean value of the iso-MoI. While the $T$=1 pairing yields
$\lambda\approx 1$, strong enhancement of the
Wigner term due to the isoscalar pairing is clearly seen
as anticipated~\cite{Wigner}.
In the SO(8) case $\lambda$ reaches the Wigner supermultiplet
limit $\lambda \sim 4$ for large $g$ (gray triangles) and drops
with decreasing $g$ as well as with increasing $T$ reaching unity
for large $T$.

In summary, we have presented the exact solution of the RG model associated to the SO(8) algebra in the context of
nuclear  n-p pairing with equal strength for the $T$=1 and $T$=0 interaction components. We have briefly discussed
a new technique for solving the Richardson equations which has a potential to become an invaluable tool in
studying integrable models for binary mesoscopic systems. The first application to the nuclear n-p pairing is
discussed from both {\it microscopic\/} as well as {\it spectroscopic\/} points of view. In particular, it is
shown that $T=0$ wave function shows alpha-like quartet structures that can be recognized by the formation of
clusters of spectral parameters containing two pair energies. At high $T$ these alpha-clusters dissolve and two
separate p-p and n-n superfluid condensates are formed. Spectroscopic consequences of these {\it microscopic\/}
processes are discussed and interpreted in terms of of generalized rotations in isospace. It is shown that the
exact solutions follow nicely the general trends predicted by the iso-cranking model.

We acknowledge fruitful discussions with S. Pittel and P. Van Isacker. This work was supported in part by the
Spanish MEC under grant No. FIS2006-12783-C03-01 and by the Polish KBN under contract No. 1~P03B~059~27. S.L.H.
acknowledges financial support from Spanish SEUI-MEC. B.E. was supported by the Spanish CE-CAM.

\end{document}